\documentclass[useAMS]{mn2e}
\usepackage{amsmath}
\usepackage{url}
\usepackage{amsfonts}
\usepackage{amsbsy}
\usepackage{subfigure}
\usepackage{verbatim}
\usepackage{amssymb}
\usepackage{amsbsy}
\usepackage{graphicx}

\renewcommand{\u}{\ensuremath{\mathbf{u}}}
\renewcommand{\d}{\ensuremath{\partial}}

\newcommand{\ex}{\ensuremath{\mathbf{e}_{x}}}
\newcommand{\ey}{\ensuremath{\mathbf{e}_{y}}}
\newcommand{\ez}{\ensuremath{\mathbf{e}_{z}}}

\newcommand{\ii}{\ensuremath{\text{i}}}

\newcommand{\ee}{\text{e}}

\title[The convective overstability]{On the convective overstability in protoplanetary discs}
\author[Henrik N. Latter]{Henrik N. Latter$^{1}$\thanks{E-mail:
    hl278@cam.ac.uk} \\
$^{1}$ DAMTP, University of Cambridge, CMS, Wilberforce Road,
Cambridge CB3 0WA, UK}

\begin{document}

\maketitle

\begin{abstract}

This paper explores the driving of low-level hydrodynamical activity
in protoplanetary-disc dead zones. 
A small adverse radial entropy
gradient, ordinarily stabilised by rotation, excites
 oscillatory convection (`convective
overstability') when thermal diffusion, or
cooling, is neither too strong nor too weak. I revisit the linear
theory of the instability, discuss its prevalence in protoplanetary
discs, and show that unstable modes are exact
nonlinear solutions in the local Boussinesq limit. Overstable modes
cannot grow indefinitely, however, as they are
subject to a secondary parametric
instability that limits their amplitudes to
relatively low levels. If parasites set the
saturation level of the ensuing turbulence then the convective
overstability
is probably too
weak to drive significant angular
momentum transport or to generate vortices. But I also discuss 
an alternative, and far more vigorous, saturation
route that generates radial `layers' or `zonal flows' 
(witnessed also in semiconvection).
Numerical simulations are required to determine which outcome is
favoured in realistic discs, and consequently how important the
instability is for disc dynamics. 
\end{abstract}


\section{Introduction}

It is increasingly clear from observations that
protoplanetary discs are highly structured --- exhibiting gaps,
asymmetries, and spirals --- and thus deviate
significantly from simple smooth disc models (Andrews et al.~2011, Muto et
al.~2012, Perez et al.~2014, Brogan et al.~2015). Besides
these observed features, theory predicts 
an annular region between roughly 1 AU and 10 AU in which the
magnetorotational instability
fails to initiate turbulence,
but where hydrodynamical processes may hold sway (Gammie 1996,
Armitage 2011). Originally termed a `dead zone' (though not nearly so
dead as first thought), this region plays a critical role in theories
of outburst behaviour and vortex and planet formation
(e.g.\ Armitage et al.~2001, Varni\'ere \& Tagger 2006, Kretke et
al.~2009, Chatterjee \& Tan 2014). 
 
One possible source of hydrodynamical activity
is convection. Irradiation from the protostar will certainly generate a
negative radial temperature gradient in the disc, though it is unclear
if this is sufficient to force the entropy $S$ to decrease outward as well (a
necessary precondition for convection).
In fact, observations indicate that on radii $\gtrsim 20$ AU most disks 
exhibit $dS/dR>0$ and hence convection is impossible 
(Andrews et al.~2009, 2010; Isella et al.~2009, Guilloteau et al.~2011). 
The situation is uncertain, however, at smaller radii, near the disk
surface, 
and around structures such as
dead zones, opacity transitions, and gaps; here the sign of the gradient may well be reversed. 
This paper is relevant to specific disc locations for which this is
indeed the case. 

The other problem that faces convection 
is the discs's strong differential rotation, which 
would easily negate its
negative entropy gradient; according to
the Solberg-H\o{}iland criterion, protoplanetary discs are stable.
However, certain double-diffusive processes have found
ways around this constraint. Examples include:
a resistive instability 
that employs diffusing magnetic fields (Latter et al.~2010a),
the subcritical baroclinic instability (SBI, Lesur \& Papaloizou 2010),
and the convective overstability (Klahr \& Hubbard 2014, Lyra 2014),
the latter two making use of thermal diffusion (or cooling). It is to the
convective overstability that this work is devoted. 
 
Thermal diffusion introduces a crucial time
lag between an inertial wave's dynamical oscillation 
(an epicycle, essentially)
and its associated thermodynamic
oscillation. After half an epicycle a fluid blob returns
to its starting radius at a different temperature than its
surroundings. As a consequence, it suffers a buoyancy acceleration that
amplifies the initial oscillation, leading to runaway growth. 
Such overstable convection was first touched on by
Chandrasekhar (1953, 1961) but, as interest originally lay in
stellar interiors, researchers focussed on
cases in which the oscillations arose not from rotation but from
magnetic tension (Cowling 1957) or composition gradients
(`semi-convection', Kato 1966).
It is only recently, decades later, that oscillatory convection has
been raised
in the context of accretion discs (Klahr \& Hubbard 2014, Lyra 2014),
even though it could play an important part in dead zone dynamics.
Indeed, the local simulations of Lyra (2014)
suggest that the instability's nonlinear saturated state transports
a respectable amount of angular momentum and even generates
vortices, possibly in conjunction with the SBI. 

This paper will revisit both the linear and nonlinear theory of the
convective overstability, remaining in the Boussinesq approximation
throughout. I reproduce analytic expressions for the growth rate,
and show that the fastest
growing mode possesses a local growth 
rate of $|N^2|/\Omega\sim 10^{-3}\Omega$, where $N$ and $\Omega$ are the radial 
buoyancy and rotation frequencies of the disc. Its
vertical wavelength is short, of order $\sqrt{\xi/\Omega}\sim 10^{-2}
H$ at 1 AU, where $\xi$ is the thermal diffusivity 
and $H$ is the disc scale height,
while its radial wavelength is much longer and connects up
to the global structure. I also discuss the prevalence of convective
overstability in realistic discs, and conclude (in agreement with Lin
\& Youdin 2015) that it is not widespread, perhaps only appearing in inner
disc regions, dead zones, or near gaps.
I also show that the unstable modes are
nonlinear solutions to the governing equations of Boussinesq
hydrodynamics, and thus can grow to arbitrarily large amplitudes, at
least in principle.  
Before a mode grows too powerful, however, it is attacked by parasitic
instabilities, the foremost of which involves the well known parametric resonance
between inertial waves and an epicycle (e.g.\ Gammie et al.~2000). 
Subsequently, an analytical estimate for the overstability's
saturation level can be
derived that predicts
a maximum amplitude of $|N^2|/\Omega^2$ over the
background. 

I discuss the nature of the 
ensuing turbulence, and consider connections with
semiconvection, as well as the SBI. 
In particular,
I argue that if the characteristic amplitude of
the turbulent state is determined by the parasitic modes 
and if $|N|\ll \Omega$, 
then its motions will be too axisymmetric and too weak to
transport appreciable angular momentum or to
generate vortices. If, on the other hand, the saturation culminates in
layer formation, as can occur in semiconvection,
the turbulent velocities may be orders of magnitude
greater and more interesting dynamics may ensue. 
Numerical simulations are needed to determine
which outcome is
realistic.

The overstability's turbulent stirring
could agitate dust grains, impeding settling
but also enhancing the collision frequency and collision speeds of
0.1-1 m particles.
Lastly, the convective overstability, being essentially global in
radius, may be connected to global dynamics and (more speculatively)
excite a small amount of eccentricity, though this
cannot be tested in the local model used here.

\section{Model equations}

Being interested in small scales and subsonic flow, I
approximate the protoplanetary disc with the Boussinesq shearing
sheet. This model describes a small `block' of disc centred at a
radius $R_0$ moving on the circular orbit prescribed by $R_0$ and at
an orbital frequency of $\Omega$. The block is represented in
Cartesian coordinates with the $x$ and $y$ directions corresponding to
the radial and azimuthal directions, respectively 
(see Goldreich \& Lynden-Bell 1965). 

The governing equations are
\begin{align} \label{GE1}
&\d_t \u + \u\cdot\nabla\u = -\frac{1}{\rho}\nabla P -2\Omega \ez\times
\u  \notag \\ 
& \hskip2cm + 2q\Omega\, x\,\ex -N^2\theta\,\ex + \nu\nabla^2\u, \\
& \d_t\theta + \u\cdot\nabla\theta = u_x + \xi\,\nabla^2\theta,\label{GE2} \\
& \nabla\cdot \u = 0, \label{GE3}
\end{align}
where $\u$ is the fluid velocity, $P$ is pressure, $\rho$ is the
(constant) background density, and $\theta$ is the buoyancy variable. The
shear parameter of the sheet is denoted by $q$, equal to $3/2$ in a
Keplerian disc, and the buoyancy frequency arising from the radial
stratification is denoted by $N$. 
Being interested
in the optically thicker dead zone, we employ thermal diffusion rather
than an optically thin cooling law, as is done in Klahr \& Hubbard
(2014) and Lyra (2014). Viscosity, denoted by $\nu$, has been included for
completeness but will be usually set to 0. 
 
Following Lesur \& Papaloizou (2010), the stratification length $\ell$ has
been absorbed into $\theta$. Thus
$\theta \propto \ell^{-1}(\rho'/\rho)$, where $\rho'$ is the perturbation to the
background density. The (squared) buoyancy frequency can be determined from
\begin{align} \label{Nsq}
N^2 = - \frac{1}{\gamma \rho}\frac{\d P}{\d R}\frac{\d \ln\left(P\rho^{-\gamma}\right)}{\d R},
\end{align}
evaluated at $R=R_0$. In the above $\gamma$ is the adiabatic index.
Another important quantity is the (squared) epicyclic frequency
\begin{equation} 
\kappa^2 = 2(2-q)\Omega^2.
\end{equation}
In addition to $q$, the system can be specified by two other
dimensionless parameters.
The Richardson number
measures the relative strength of the radial stratification;
it is denoted by $n^2$ and defined via
\begin{equation}
n^2 = -\frac{N^2}{\kappa^2}.
\end{equation}
In thin astrophysical discs, $n^2$ is generally small (see Section 3.4).
The Prandtl number helps quantify the separation in scales between the
thermal lengthscale and the viscous lengthscale; it is denoted by Pr
and defined via
\begin{equation}
\text{Pr} = \frac{\nu}{\xi},
\end{equation}
it too is generally small. Finally, though the outer scale does not
appear in the governing equations, it can be useful to define 
the
Peclet number
\begin{equation}
\text{Pe}= \frac{H^2\kappa}{\xi},
\end{equation}
where $H$ is the vertical scale height. In our problem, 
this parameter helps quantify
the separation in scales between the instability length and the
disc thickness.

\section{Linear theory}

In this section I revisit the analyses presented in Klahr \& Hubbard
(2014) and Lyra (2014), and provide an analytical expression for the growth rate in
the limit of small Richardson number. I explicate
the physical mechanism of instability and apply these results to the
conditions expected in protoplanetary discs.

\subsection{Inviscid eigenproblem}

Equations \eqref{GE1}-\eqref{GE3} yield the equilibrium,
$\u=- q\Omega x\,\ey$, $\theta=0$, and $P$ a constant. 
This is perturbed by disturbances
proportional to $\ee^{st + \ii Kz}$, where $s$ is a
(complex) growth rate and $K$ is a (real) vertical wavenumber. We assume that
any radial variation exhibited by our modes lie on scales larger than
the shearing sheet and does not interfere with its local
physics. Viscosity is also dropped, to ease the analysis.

Denoting perturbation with primes, the linearised equations are
\begin{align} \label{lin1}
& s\,u_x' = 2\Omega\, u_y' - N^2 \theta', \\
& s\,u_y' = (q-2)\Omega\, u_x', \label{lin2}\\
& s\,\theta'= u_x' - \xi k^2\,\theta'.\label{lin3}
\end{align}
Our ansatz ensures that $u_z'=0$, via the incompressibility condition,
and $P'=0$, via the $z$-component of the momentum equation. The
dispersion relation for these modes is easily obtained:
\begin{align} \label{disp}
s^3 + \beta s^2 + (N^2+\kappa^2)s + \beta \kappa^2=0, 
\end{align}
where $\beta= \xi\,K^2$ is the (length-scale dependent) cooling rate. 
Apart from differences in notation, Eq.~\eqref{disp} agrees with Eq.~(18)
in Klahr \& Hubbard (2014), Eq.~(21)
in Lyra (2014), and the inviscid version of Eq.~(B2) in Guilet \&
M\"uller (2015).

When there is no thermal diffusion whatsoever Eq.~\eqref{disp} is easy
to solve and one obtains buoyancy-assisted epicyclic
motion. Instability occurs when
\begin{equation}
N^2 + \kappa^2 < 0,
\end{equation}
i.e.\ the Solberg-H\o{}iland criterion.
When $\xi$ is nonzero the dispersion relation is a cubic and the analytic
solution messy. A numerical solution, using 
fiducial parameters, is plotted in
Fig.~1.

In the
natural limit of small Richardson number $n^2 \ll 1$, 
a convenient analytical expression for $s$
is available, as Guilet \& M\"uller (2015) show. They set $s= s_0 + s_1
n^2+\dots$ and to leading order Eq.~\eqref{disp} becomes
$$ (s_0^2+\kappa^2)(s_0+\beta) =0,$$
which yields 
a decaying energy mode and two epicycles. We take one of the epicycles,
$s_0=\ii \kappa$,
and at the next order obtain
$$ s_1 = -\frac{1}{2}\frac{\kappa^2}{\kappa^2+\beta^2}\left(\beta-\ii\kappa\right) .$$
The real part of the growth rate is hence
\begin{equation}\label{sapprox}
\text{Re}(s) = -\frac{1}{2}\frac{\beta\,N^2}{\kappa^2+\beta^2} +\mathcal{O}(n^4\kappa),
\end{equation}
which matches the full solution to Eq.~\eqref{disp} for all $k$ (or
$\beta$) in the appropriate limit.
Maximum growth is achieved when $\beta=\kappa$, i.e. when
$K=\sqrt{\xi/\kappa}$, for which
\begin{equation}
\text{Re}(s_\text{max}) = -\frac{1}{4}\frac{N^2}{\kappa},
\end{equation}
which agrees with Eq.~(32) in Lyra (2014). 

Three things are noteworthy.
First, the Solberg-H\o{}iland criterion no longer determines the onset of
instability. Instead, growth occurs when
$N^2 <0$
(the Schwarzchild criterion); thus the convective overstability has used
thermal diffusion to negate the stabilising influence of rotation.
Second, the maximum growth rate
is independent of the magnitude of thermal diffusion, even though 
it is crucial to the existence of the instability. Thermal diffusion
works as a catalyst of instability, 
but only fixes its characteristic lengthscale. 
Third, because its radial wavenumber is zero the mode's group speed
$\d\text{Im}(s)/\d K$ is small, proportional to $n^2$ (and hence much
less than the phase speed). As a consequence, 
there is no risk that the mode energy
propagates far away before it grows to appreciable amplitudes. 

\begin{figure}
\begin{center}
\scalebox{0.6}{\includegraphics{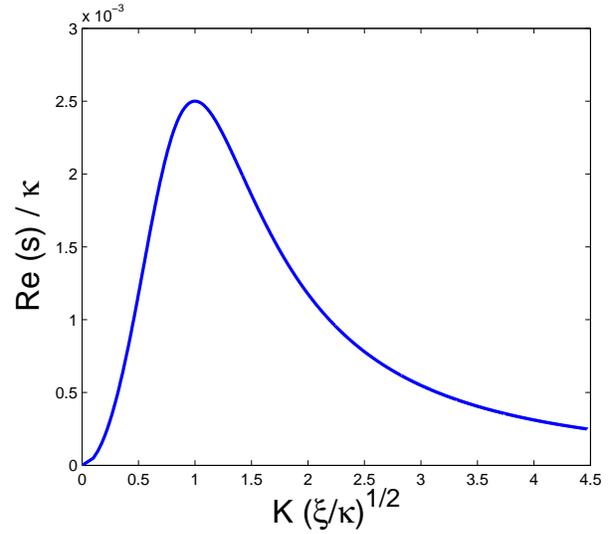}}
 \caption{Real part of the convective overstability's growth rate as a
   function of wavenumber $K$. The Richardson number is
   $n^2=0.01$. The solid line represents the full solution to
   Eq.\eqref{disp}, but the analytic approximation of
   Eq.~\eqref{sapprox} 
  is indistinguishable.} \label{DispReln}
\end{center}
\end{figure}

The eigenfunction of the unstable mode itself
consists of a vertical stack of planar fluid sheets
each undergoing slowly growing epicycles and each
communicating with its neighbours via thermal diffusion.
Because vertical motion is absent from the mode, it is relatively
impervious to the disc's vertical structure, in particular a stable
stratification.  
Note that if a
scale-free cooling law is adopted, rather than thermal diffusion
(Klahr \& Hubbard 2014, Lyra 2014),
the unstable modes can possess an arbitrary dependence on $z$. In
fact, such a model predicts that
\emph{every} scale, from the viscous cut-off all the way to $H$, grows
at the same maximum rate, a situation that is somewhat artificial and
may pose problems when the nonlinear dynamics are simulated.

Putting to one side its vertical dependence, the convective
overstability
can also
be identified as
the local manifestation of a global eccentric mode, with azimuthal
wavenumber equal to 1 (Ogilvie 2001, Ogilvie \& Barker 2014).
The convective overstability may thus potentially excite
the disc's eccentricity, though this is an idea not pursued in this
paper.

Lastly, there also exist modes
with non-zero radial wavenumbers that grow slower and are not
treated here (see Lyra 2014). Non-axisymmetric modes, on the other
hand, offer
only a short period of transient growth before being sheared out, and
are also neglected.

\subsection{Influence of viscosity}

In the regime $n^2\gg \text{Pr}$, viscosity only adds a small
correction to the maximum growth rate. But it does damp
small scales that would be otherwise unstable. In Appendix A, I calculate 
the critical wavenumber
upon which short modes are stabilised. For small Prandtl number it may be
estimated by
\begin{equation} 
K_c \approx \left(\frac{n^2}{\text{Pr}}\right)^{1/4}K_\text{fast},
\end{equation}
where $K_\text{fast}$ is the wavenumber of fastest growth. Note that
the $1/4$ power means that the spatial separation between the fastest
growing and marginal modes is not as vast as one would first think. 

We may also derive a revised stability condition for
the convective overstability. When viscosity is present the Schwarzchild
criterion is replaced by
\begin{equation}\label{critter}
N^2 < -\frac{2\text{Pr}}{1+\text{Pr}}\Omega^2.
\end{equation}
The entropy gradient must be negative \emph{and} sufficiently
strong. However, owing to
the smallness of Pr in real discs, the criterion is
only slightly modified. On the other hand, 
in numerical simulations where $\nu$ is
unrealistically large, Eq.~\eqref{critter} should be kept in mind.

\subsection{Instability mechanism}

Figure 2 illustrates the basic mechanism of the convective
instability, drawing on arguments in Cowling (1958). The mechanism is
almost identical to that driving the SBI, as described in Lesur \&
Papaloizou (2010). The only difference is that the convective
overstability forces fluid blobs to execute
epicycles, while the SBI forces them to circulate around a vortex.

Consider a local
patch in the disc-plane exhibiting a weak entropy gradient so
that `colder' fluid is at larger radii, and `hotter' fluid is at
smaller radii. Suppose a fluid blob at a given radius, associated with
an entropy of $S_0$, is made to undergo epicyclic motion, panels (a)
and (b). 

\begin{figure}
\begin{center}
\scalebox{0.4}{\includegraphics{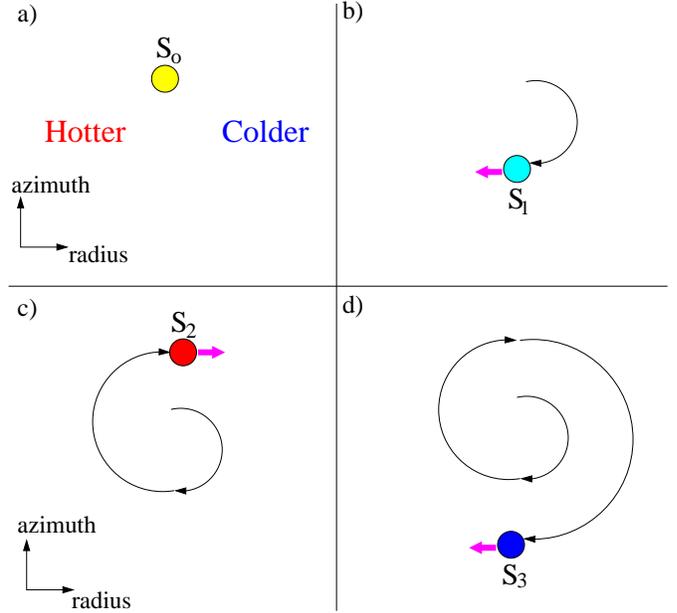}}
 \caption{Four panels indicating the convective overstability
   mechanism. In panel (a) a fluid blob is embedded in a radial
   entropy gradient. In panel (b) it
    undergoes half an epicycle and returns to its original radius with
  a smaller entropy than when it begun $S_1<S_0$. It hence feels a
  buoyancy acceleration inwards and the epicycle is amplified. The
  process occurs in reverse once the epicycle is complete, shown in
  panel (c), where now $S_2>S_0$. The
  oscillations hence grow larger and larger.}\label{Cartoon}
\end{center}
\end{figure}

Panel (b) shows the blob after half an epicycle, when it has returned to its
original radius. During its outward excursion it has come into contact
with colder fluid and has thus exchanged some of its initial heat via
thermal diffusion. As a consequence, when it returns to
its starting radius it is cooler and possesses a new entropy $S_1<
S_0$. Because of this entropy difference the blob suffers an inward
buoyancy force (represented by the magenta arrow) that boosts the
amplitude of the epicycle. 

Panel (c) shows the blob after executing a full epicycle. Again it is
back at its starting radius but now it has greater entropy than its
surroundings $S_2>S_0$ because it has attempted to equilibriate with
the hotter fluid at smaller radii. The blob now feels an outwardly
directed buoyancy force that further amplifies the epicycle. And panel
(d) shows the next phase, where the process runs away.

Instability would be quenched if thermal diffusion was too efficient
or too inefficient. In the first (isothermal) case, at every stage of the epicycle
the fluid blob would possess the same entropy as its surrounding. It
would hence never feel a buoyancy acceleration. In the second
(adiabatic) case, the fluid blob's entropy would never deviate
 from $S_0$ and it would execute buoyancy-adjusted epicycles.

\subsection{Parameter values and physical scales}

\subsubsection{Buoyancy frequency 
and characteristic timescales}

I first discuss the sign and magnitude of $N^2$. This quantity must be negative
for there to be instability, but what do recent observations have to
say about this? Let us examine the stability of the disc midplane first.
Assuming that $\gamma = \frac{7}{5}$, the inviscid 
instability condition may be re-expressed as
$q_\rho > \frac{5}{2}q_T$, where $q_\rho=d\ln \rho/d \ln R$ and
$q_T=d\ln T/d \ln R$ with $T$ temperature. 
Taking $\rho$ at the midplane,
this becomes
\begin{equation}\label{instcrit}
q_\Sigma > \tfrac{5}{2}q_T + q_H,
\end{equation}
where $q_\Sigma = d\ln \Sigma/d\ln R$ and $q_H=
d\ln H/d \ln R$, and in which $\Sigma$ is surface density. 
Equation \eqref{instcrit}
basically states that instability favours disc radii with fairly flat density
profiles, and discs that are less flared. In addition, the stronger the
negative temperature gradient the more likely instability, as expected.

To date, the various $q$ parameters
have been estimated from (sub-)mm observations of
some two dozen pre-main sequence stars, mainly in the
Taurus and Ophiucus star-forming regions (Andrews et al.~2009, Isella et
al.~2009, Guilloteau et al.~2011). 
Generally $q_T$ lies between -0.6 and -0.5 and $q_H= 1.04-1.26$.
There is greater variation in the density structure. Andrews et
al.~(2009) find that $q_\Sigma$ can fall between $-1$ and $-0.4$. As a
consequence, all but one of the discs in their sample fail to satisfy
Eq.~\eqref{instcrit}, with AS 209 perhaps marginally unstable. 
Isella et al.~(2009) and Guilloteau et al.~(2011) 
obtain a larger spread, with $q_\Sigma$ varying between
$-1.5$ and $0$ on intermediate to long radii, far from the disk inner
edge.  Though
some inner regions may satisfy Eq.~\eqref{instcrit}, most of the discs
in this sample exhibit an insufficiently flat density structure and are hence
also generally stable\footnote{It should be remembered that, because
  instabilities tend to erase the unstable conditions from which they
  erise, the observations cannot show disc structures that are `about
  to be attacked' by instability but rather 
  structures after the instability has had its way.}.

The conditions for convective overstability improve, however, the
further from the disc midplane. A locally isothermal model with
realistic power
laws in density and temperature reveals that the magnitude of
$q_\rho$ has decreased by
$\sim 30\%$ at $z=(3/4)H$. This is sufficient to push some discs to
marginal stability or perhaps better, 
but a more thorough study of disc structure
is necessary to settle the issue. 
Note that locations higher up in the disc, $z>(3/4)H$,
are probably inappropriate venues for overstability on
account of the magnetorotational turbulence, wind launching, and/or
associated planar jets 
(Fleming \& Stone 2003, Bai \& Stone 2013, Gressel et al.~2015).

Taking these results on face value, we conclude (as do Lin \& Youdin
2015) that most locations in most protoplanetary discs exhibit a positive
$N^2$ and are hence stable to the convective overstability (as well
as the SBI). 
 At smaller radii ($\sim 1$ AU) the picture is less clear
because there the disk structure is less well constrained by the
observations. 
This is also true in and around
conspicuous disc features such as dead zones (which may be partly
shadowed by the disc's hotter inner radii), opacity transitions,
and edges, because the
fitting models assume smooth disc profiles. These regions could in principle
possess unstable entropy gradients, but more advanced numerical
modelling is needed to establish whether this is actually the case 
(see Faure et al.~2014, Flock et al.~2015). Finally, because gas
off the midplane possess weaker radial density gradients, instability
may also favour locations higher up in the disk. 
 The working hypothesis of
this paper is that there are indeed certain disc regions, in particular
dead-zones, that are convectively overstable.

Supposing that a disc region is overstable, 
in order for the instability to have a measurable effect,
it must grow sufficiently quickly.
The maximum growth rate depends closely on
the strength of the magnitude of $N^2$, via Eq.~\eqref{sapprox}. 
But, as discussed above, it is unclear what values it should take.
The best we can do is  assume that both the
pressure and entropy decrease with radius so that $\d_R \sim 1/\ell$. 
Then
\begin{equation}
N^2 \sim \left(\frac{H}{\ell}\right)^2 \Omega^2,
\end{equation}
from Eq.~\eqref{Nsq}, 
where the stratification length $\ell\lesssim R$ for a smooth gradient (as might be the case deeper
in a dead zone), 
and $\ell \gtrsim H$ near an abrupt structure (a dead zone edge or gap edge,
for example). Throughout the paper we take a conservative approach and
consider $\ell \sim R$. With the standard scaling $H/R \sim 0.05$ we obtain,
\begin{equation}\label{est}
\text{Re}(s) \sim 10^{-3}\,\Omega.
\end{equation}
Thus the efolding time at 1 AU is roughly 1000 years, plenty of scope
for the instability to develop within a protoplanetary disc's
lifespan. It should not be forgotten that Eq.~\eqref{est} may be an
underestimate near more abrupt disc structures, for which $\ell\gtrsim H$.

\subsubsection{Prandtl number and characteristic lengthscales}

The previous subsection assumes that the fastest growing scales fit into the
disc or are not so small that viscosity stabilises them. This needs to
be checked. From
\eqref{sapprox} the dominant vertical scale is $K\sim \sqrt{\Omega/\xi}$. A
standard expression for the thermal diffusivity at the midplane at 1 AU is
\begin{align}\notag
&\xi = 2.39\times 10^{12}\left(\frac{\rho}{10^{-9}
    \text{g}\text{cm}^{-3}}\right)^{-2}\left(\frac{T}{100\,\text{K}}\right)^3
 \\ & \hskip3cm
\times \left(\frac{\kappa_\text{op}}
{\text{cm}^2\text{s}^{-1}}\right)^{-1}\,\text{cm}^2\text{s}^{-1},
\end{align}
where $\kappa_\text{op}$ is opacity. The
reference values are drawn from the minimum mass solar nebula at a
few AU (Hayashi et al.~1985) and calculated opacities at low
temperatures (Henning \& Stognienko 1996). Hence
the dominant vertical wavelength at 1 AU is
\begin{equation}
\lambda \sim K^{-1}\sim  10^{10} \text{cm} \sim 0.01\,H,
\end{equation}
which fits comfortably into the disc. It follows that the Peclet number is
$\sim 10^4$. 

The convective overstability exhibits 
relatively small vertical scales, endorsing the
adoption of the local Boussinesq model, yet still far larger than the
viscous length. At 1 AU, $\nu \sim 10^5$ cm$^2$ s$^{-1}$ and the
Prandtl number is $\sim 10^{-7}$. 
On the other hand, the radial scales are unconstrained by the analysis
(similarly to magnetorotational channel modes, Balbus \& Hawley 1991),
and must lie on scales of order $\ell$ or $R$.

\subsection{Unstable modes are nonlinear solutions}

The final important point is that the linear modes explored in this
section are also exact nonlinear solutions to the governing equations.
Both $\u$ and $\theta$ depend on $z$ and $t$, but $u_z'=0$. Therefore
all the nonlinear terms in Eqs \eqref{GE1}-\eqref{GE2} vanish:
$$\u'\cdot\nabla\u'= \u'\cdot\nabla\theta'=0.$$
As a consequence, a convectively overstable mode
will grow exponentially even after it leaves the linear regime, and in
theory can achieve arbitarily large amplitudes. A similar property is
shared by magnetorotational channel flows (Goodman \& Xu 1994).

Of course, the exponential growth cannot continue indefinitely. For a
start, the system will ultimately violate the Boussinesq assumptions, i.e.\
subsonic flow and small thermodynamic variation. It may also be that
the mode's global radial structure intervenes to halt the
runaway. Perhaps, before either comes into play, parasitic modes,
feeding on the mode's strong shear, destroy the mode and
initiate a period of hydrodynamical turbulence. It is this last
possibility that we explore next.

\section{Parasitic instabilities}
\subsection{Linearised equations}

In this section, we view the growing oscillations of
Section 3 as part of the basic state and subsequently
 explore growing perturbations to this state. 
 It is assumed the
 oscillations have reached an amplitude characterised by the shear
 rate $S$, and though we permit $S\sim\Omega$ or larger, 
to ease the calculations
the buoyancy frequency is assumed small, so that $|N| \ll
\kappa,\,S$. Because the parasitic growth rates will be $\sigma\sim
S$, the buoyancy term may then be dropped 
from the perturbation equations,
and the thermal equation decouples. 
We may also omit the slow growth of the convectively overstable
mode itself, as it is negligible compared to $\kappa$ and $S$.
Finally, we set $q=3/2$, and thus $\kappa=\Omega$. The equilibrium
to leading order may now be written as
$$ u_x = S \cos \Omega t\,\cos K z, \qquad u_y= -\tfrac{3}{2}\Omega\,x
-\tfrac{1}{2}S \sin \Omega t \cos K z, $$
where we have used the eigenvector of Eqs \eqref{lin1}-\eqref{lin2}
to describe the oscillatory component of the equilibrium.

Units are chosen so that $K=1$, $\kappa=1$, and $\rho=1$. The background state is
disturbed by velocity and pressure perturbations taking the standard
Floquet form, 
$$ \hat{\u}(t,z) \ee^{\sigma t + \ii m z+ \ii k x},
 \qquad \hat{p}(t,z) \ee^{\sigma t + \ii m z+ \ii k x},$$
where the hatted variables are $2\pi$-periodic in both $t$ and $z$. 
Here $k$ is a radial wavenumber, 
and $\sigma$ and $m$ are Floquet exponents; the former serves as the
growth rate of the perturbation. Note that I neglect non-axisymmetric
disturbances; because such modes will be sheared out quickly, they are less
effective at killing their hosts. Only at very large $S$ will they be important.  

The linearised equations governing the
evolution of these perturbations are
\begin{align} 
\sigma\hat{u}_x &= -\d_t \hat{u}_x -\ii k \,\epsilon U\,\hat{u}_x -\epsilon \d_zU\,\hat{u}_z -
\ii k \hat{p} + 2 \hat{u}_y, \label{LIN1}\\
\sigma\hat{u}_y &= -\d_t \hat{u}_y -\ii k \,\epsilon U\,\hat{u}_y - \epsilon
\d_z V\,\hat{u}_z
-\tfrac{1}{2}\hat{u}_y, \\
\sigma\hat{u}_z &= -\d_t \hat{u}_z -\ii k \,\epsilon U\,\hat{u}_z - \d_z\hat{p}-\ii m
\hat{p}, \\
0 &= \ii k \hat{u}_x +  \d_z \hat{u}_z+ \ii m \hat{u}_z , \label{LIN2}
\end{align}
where the background convective oscillation is represented by
\begin{align}\label{backg}
U = \cos t\,\cos z, \qquad V= -\tfrac{1}{2}\sin t \cos z,
\end{align}
and
the parameter $\epsilon= S/\Omega$ measures the amplitude of the
background convective oscillation. 

We now have a two-dimensional eigenvalue problem in both $t$ and
$z$. The eigenvalue is $\sigma$, while the governing parameters are
simply $\epsilon$ and $k$. Generally Eqs \eqref{LIN1}-\eqref{LIN2}
must be solved numerically, but in the limits of small and large $S$
some analytical progress can be made. We treat these asymptotic limits
first and then give the numerical solutions.

\subsection{Asymptotic solutions}
\subsubsection{Large amplitude oscillations}

The first limit is perhaps the easiest to understand, though not the
most relevant. I assume that
the background oscillations are extremely strong, with shear rates
 $S\gg \Omega$.  
In this regime, parasites grow so fast 
that the oscillation is destroyed before
completing even one cycle. As a consequence, 
it should be
regarded as `frozen' and, furthermore, 
the background differential rotation omitted.
The problem then reduces to determining the
stability of a spatially periodic shear called Kolmogorov
flow (Meshalkin \& Sinai 1961), an especially well-studied
model problem for which numerous results have been proven 
(see, for example, Beaumont 1981 and Gotoh et al.~1983). 

The inflexion
point theorem suggests the flow is unstable, and indeed Drazin \&
Howard (1962) show that instability occurs on $k<1$. A
reasonable approximation to the growth rates is derived by Green
(1974), who uses a truncated Fourier series to obtain
\begin{equation}\label{largeass}
\sigma^2 \approx  \frac{k^2(1-k^2)}{2(1+k^2)}\,\epsilon^2,
\end{equation}
when $m=0$. Here $k$ should be understood as the wavenumber in the
direction of the shear at any given instant.
In agreement with our initial assumption, Eq.~\eqref{largeass} yields 
a growth rate $ \sim S$.

\subsubsection{Small amplitude oscillations}

Before such large amplitudes are achieved the oscillation will be
destroyed by a different class of parasitic mode involving a
parametric 
resonance between the growing epicycle and two inertial waves. The 
instability is a relation of the famous elliptical instability, which 
disrupts
vortices both in and outside of protoplanetary discs (Pierrehumbert
1986, Bayly 1986, Kerswell 2002,
Lesur \& Papaloizou 2009, Railton \& Papaloizou 2014). A variant of
the instability also attacks accretion discs themselves when the
streamlines deviate from non-circular orbits (Goodman 1993), as in the
case of eccentric (Papaloizou 2005, Barker \&
Ogilvie 2014) and warped discs (Gammie et al.~2000, Ogilvie \& Latter
2013). Locally this deviation appears as an oscillation with
frequency equal to $\kappa$ and similar in form to the convective
overstability mode.

The parametric instability can be understood as a special case of a
three-wave coupling (Gammie et al.~2000). The primary overstable oscillation,
with frequency $\kappa$,
provides a means by which two linear inertial waves, of frequencies
$\omega_1$ and $\omega_2$, can work together
to draw out energy from the primary. In order
for this to happen a
resonance condition $\omega_1+\omega_2 = \kappa$
must be met. Resonance only occurs for a discrete set of $k$,
each corresponding to a different vertical mode number $n$.

\begin{figure}
\begin{center}
\scalebox{0.55}{\includegraphics{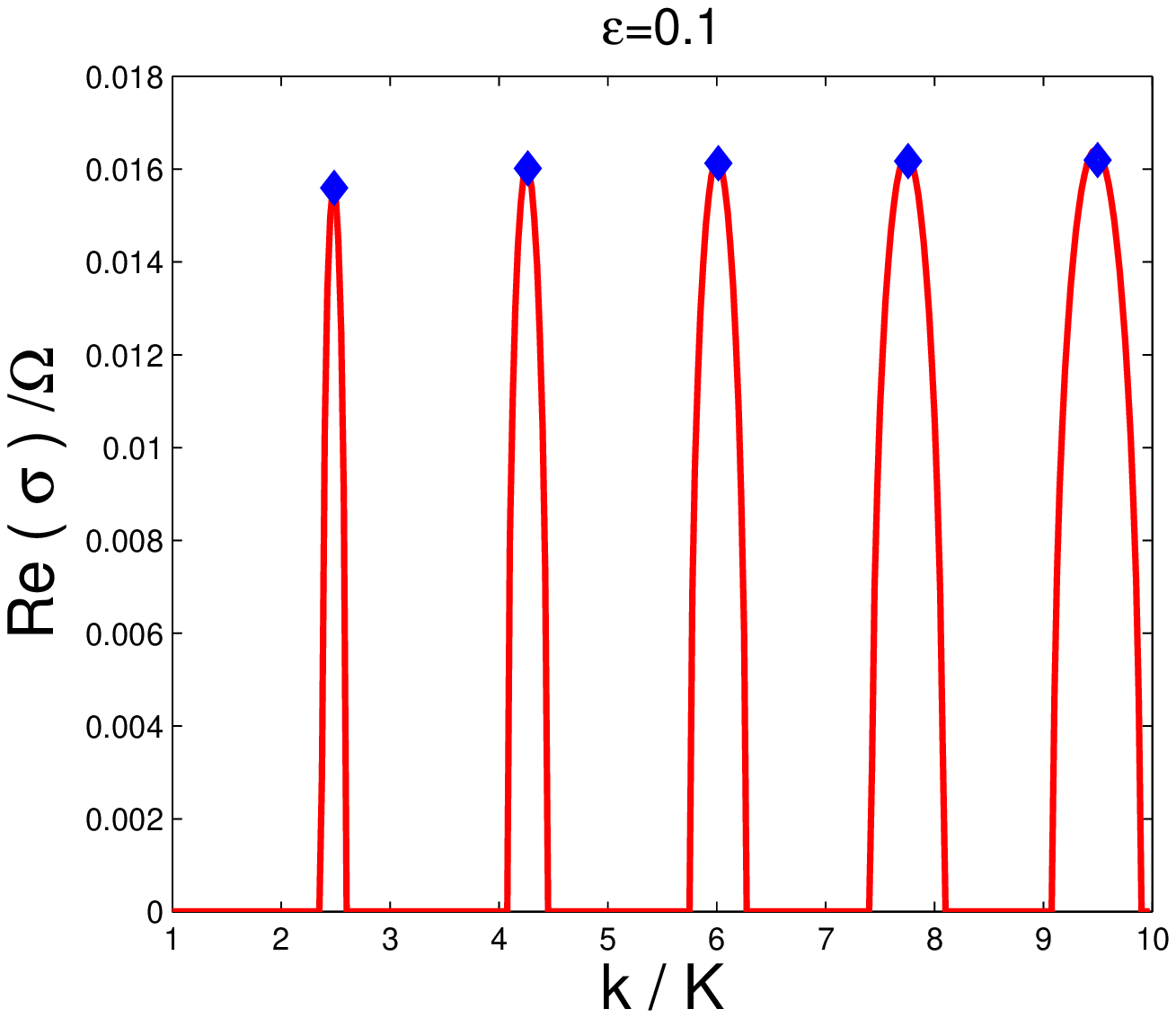}}
\scalebox{0.55}{\includegraphics{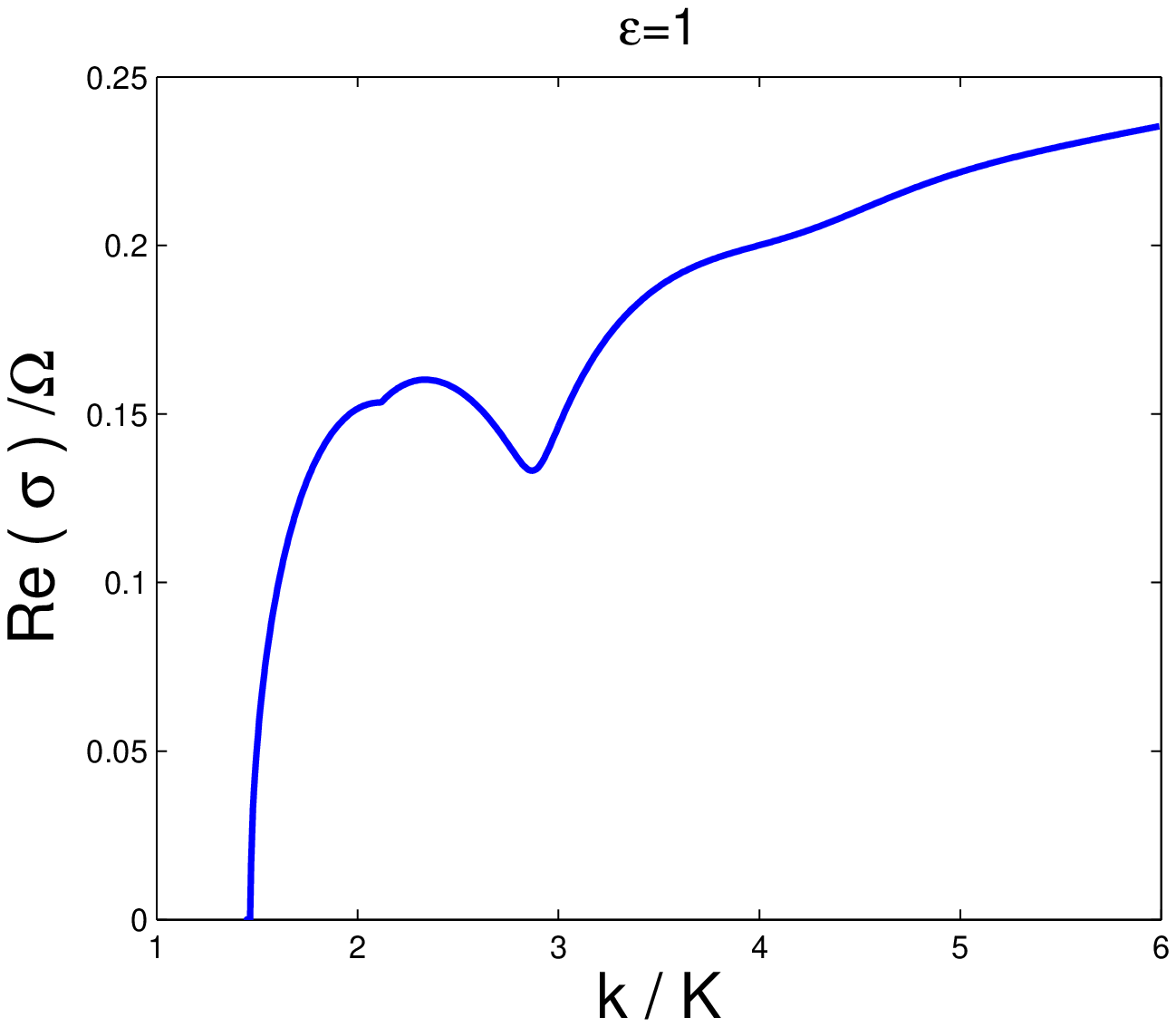}}
 \caption{Parasitic growth rates $\sigma$ as a function of radial
   wavenumber $k$ for two different amplitudes. The top panel is for
  $\epsilon=0.1$, the bottom for $\epsilon=1$.
   In the top panel the predictions of the asymptotic theory are
   represented by blue diamonds.}\label{Parasites1}
\end{center}
\end{figure}

\begin{figure*}
\begin{center}
\scalebox{0.55}{\includegraphics{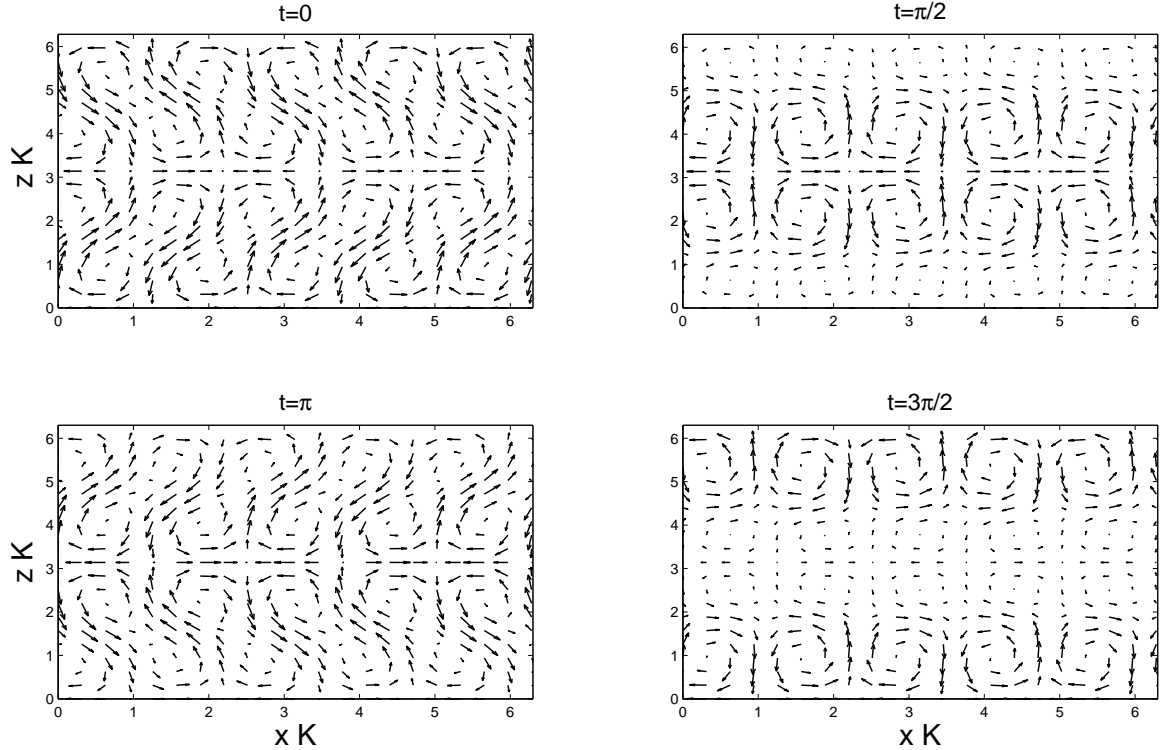}}
 \caption{Four panels showing the eigenfunction (flows in the $x$-$z$ plane)
   at four different moments during the oscillation. See Section 4.3.}
 \label{Parasites2}
\end{center}
\end{figure*}

Appendix B outlines an asymptotic theory of this resonance in
the limit of small oscillation amplitude and for $m=0$. The appropriate regime is
$|N|\ll S \ll \Omega$: thus the thermal dynamics are omitted, but the
rotational dynamics are not. 
In dimensionless variables the
resonance condition is
\begin{equation}\label{resonant}
\frac{n}{\sqrt{n^2+k^2}}+\frac{n+1}{\sqrt{(n+1)^2+k^2}} = 1,
\end{equation}
where $n$ is an integer, describing the vertical wavenumber of the
first inertial mode. The first few resonances occur at 
$k\approx 2.49,\,4.26,\,6.02,\,7.76,\,9.50$. Note that parametric
instability favours much shorter vertical scales than the shear
instability of Section 4.2.1.
The growth rate to leading order in $\epsilon$ is
given by
\begin{equation}\label{smallass}
\sigma^2=
\frac{k^2(1+2\omega_1)^2(\omega_1^2+2\omega_1-n)(\omega_1^2+n)}
{64\,n(1+n)\,\omega_1(1+\omega_1)}\,\epsilon^2,
\end{equation}
where the frequency of the first inertial wave is taken to be
\begin{equation}
\omega_1= -\frac{n}{\sqrt{n^2+k^2}}.
\end{equation}
The growth rates corresponding to the first five $n$ are plotted in
Fig.~3a, alongside the full numerical solution.
Finally, as $n\to \infty$ the growth rates plateau to a constant maximum
value, given by 
$\sigma \approx \frac{3\sqrt{3}}{32}\,\epsilon$.

\subsection{Numerical solutions}

In this section Eqs \eqref{LIN1}-\eqref{LIN2} are solved numerically
using a pseudo-spectral technique. I partition the $2\pi$-periodic $t$ and $z$ 
domains into $M_t$ and $M_z$ cells and represent each dependent variable
as a vector of length $M_t M_z$. Derivatives are described using
appropriate matrices (Boyd 2002), upon which
\eqref{LIN1}-\eqref{LIN2} may be approximated by a $4M_tM_z\times 4M_t
M_z$
algebraic eigenvalue problem, 
the eigenvalues of which are the growth rates $\sigma$. These are obtained
by the QZ algorithm or an Arnoldi method (Golub \& van Loan 1996). 
For moderate $k$, $M_t=M_z=30$ yields converged growth rates. 

The real part of the growth rate $\sigma$
is plotted in Fig.~3 as a function of
radial wavenumber $k$. Two illustrative values of the oscillation
amplitude are chosen, $\epsilon = 0.1$ and $\epsilon=1$. The vertical
wavenumber of the mode's envelope $m$ is set to zero.
The top panel lies in the small amplitude regime, and hence I also
plot the growth rates produced by the asymptotic theory of Section
4.2.2, as blue diamonds. Instability occurs in distinct
bands located at the resonant $k$ values predicted by
Eq.~\eqref{resonant}. The asymptotic growth rates are in good
agreement with the numerical results, which validates both the
analytic theory and the computations. As is typically the case, the
larger the $k$ the wider each resonant band, and at some large value the
bands overlap and the instability becomes more complicated in nature
than the simple three-wave resonance idea. This is also true the
larger $\epsilon$, as is clear from the bottom panel of Fig.~3. Here
$\epsilon=1$ and instability occurs for all $k$ above a critical
value. As expected, the growth rates are an order of magnitude greater
than the $\epsilon=0.1$ case. 
  
In Fig.~4 a representative eigenmode is shown, when $\epsilon=0.1$
and $k=2.486$. The figure comprises four snapshots in the $x-z$ plane
of the velocity taken at equally spaced moments during
its $2\pi$ cycle. The mode comes from the first band of parametric
instability in Fig.~3a, and consists of $n=1$ and $n=2$ inertial
waves coupled via the overstable oscillation. Note its characteristic
oblique motions: the radial and vertical mode speeds are most closely
correlated in panels 1 and 3 ($t=0,\,\pi$), when the background
radial shear take its largest values (cf.\ Eq.~\eqref{backg}).
The associated Reynolds stresses of the mode are thus able to
 extract the overstability's energy. Similar behaviour is observed in
 the instability of a warp (Ogilvie \& Latter 2013).

\subsection{Maximum amplitude}

It is possible to estimate a maximum saturation amplitude of the
convective overstability by comparing
the growth of the overstability itself with that of its parasites. 
Similar calculations have been carried out in the magnetorotational channel
context (Pessah \& Goodman 2009, Latter et al.~2010b), but the
overstability problem is made easier by the axisymmetry of the
parasitic modes. They cannot be sheared away by the differential
rotation.

In the previous subsections, the slow growth of the oscillation was
neglected and so $\epsilon$ was taken to be a constant.
In this section its time dependence is reinstated, so that
(in dimensionless variables)
$\epsilon = \epsilon_0 \ee^{st}$, where $\epsilon_0$ is the
oscillation's starting amplitude (the level of the background fluctuations),
and $s\sim n^2$ is its growth rate.

A crude estimate for the time it takes a parasitic mode to
overrun its host may be derived by equating the growth rates of parasite and
host, $s\sim \sigma$ (Pessah \& Goodman 2009).
The oscillation's amplitude at this point is then easy to calculate:
$\epsilon_\text{max} \sim n^2$. Returning to dimensional
units, the maximum shear is
\begin{equation} \label{saturated}
S_\text{max} \sim \frac{|N^2|}{\Omega}.
\end{equation}
Given the smallness of $N^2$ in protoplanetary discs, this is not a
large value at all, only some $10^{-3}$ the background shear rate (cf.\
Section 3.4).  In realistic discs, the convective
overstability grows so slowly that it is overrun by parasitic
modes before it extracts appreciable energy from the
thermal gradient.

This rough estimate may be improved upon by 
setting the \emph{amplitudes} of
the oscillation and parasite to be equal, rather than the growth
rates.  The parasitic mode's amplitude
 we denote by $p$, and its growth is determined from the ODE
$dp/dt = \sigma(t)p$, where $\sigma\sim \epsilon$. 
This equation yields
$ p \sim p_0 \text{exp}[(\text{exp}(st)-1)\epsilon_0/s]$, where $p_0$ is
the initial amplitude of the parasite. Setting $\epsilon=p$ produces
a nonlinear equation for the time of destruction, which may be solved
in terms of special functions. The maximum oscillation
amplitude may then be computed:
\begin{equation}
\epsilon_\text{max} \sim -n^2 W_{-1}\left(-\frac{p_0}{n^2}
\text{e}^{-\epsilon_0/n^2}\right),
\end{equation}
where $W_{-1}$ is the second real branch of the Lambert W-function
(Corless et al.~1996). Note that the final amplitude depends not only
on $n^2$ but also on the initial conditions. If next we assume that 
 both the overstability and the parasite grow
 simultaneously from the same reservoir of small amplitude noise $(< n^2)$, 
then we have the asymptotic estimate
\begin{equation}
\epsilon_\text{max} \sim n^2 \ln\left(\frac{n^2}{\epsilon_0}\right).
\end{equation}
This shows that the
dependence on the initial condition is weak and
that \eqref{saturated} is largely unaltered: convective
oscillations are destroyed at low amplitudes.

\section{Nonlinear saturation}

\subsection{Parasitic theory}

Once an isolated overstable mode is overrun by a parasite the flow
breaks down into a disordered state. Let us suppose that the
nonlinear dynamics that follow are controlled by the emergence and
decline of these fastest growing overstable modes. How might this
dynamical situation work? One possibility is that
the system settles on a weakly nonlinear
state, with the three resonant waves
joined by a small number of shorter
wavelength modes. These shorter modes, ordinarily stabilised by viscosity, will
remove the energy input by the instability and let the system
reach a statistically steady state. The nonlinear standing waves 
observed in some semiconvection simulations may be an example of this
(Mirouh et al.~2012).
Though likely in
simulations (with their relatively large Pr), 
real discs may struggle, however, to host such low-order nonlinear
dynamics. Another possibility is that the emergence and destruction of
overstable modes could control developed turbulence in which
many more modes participate. Pessah \& Goodman (2009) make such an argument to explain
the level of magnetorotational turbulence in shearing boxes, 
but being axisymmetric the parasites
considered here are much more effective because they do not shear
out (see also Latter et al.~2010b). 
Whatever its details, a parasitic theory of saturation 
would then argue that
Eq.~\eqref{saturated} sets not only the maximum amplitude of a single
overstable oscillation but also the saturation amplitude of the ensuing
turbulence. 

Equation \eqref{saturated} predicts that
the convective overstability generates only a very mild level of
turbulence in realistic discs. 
Because the typical strain rate is usually much less than $\Omega$, the
flow will be essentially axisymmetric. And if one
estimates the typical turbulent lengthscale by $1/K$, then associated
velocities will be gentle:
\begin{equation}\label{turb}
v_\text{turb}\sim 10^{-2}\,n^2\,c_s
\end{equation}
at 1 AU, where $c_s$ is the local sound speed (cf.\ Section 3.2).
Because of its low amplitude and axisymmetry, the saturated state 
should not drive
significant angular momentum transport nor
generate vortices via SBI (or another mechanism). The appearance of
both phenomena in the simulations of Lyra (2014) can be attributed to
a strong stratification, $n^2 \sim 0.1$, which may occur only in
special regions of the disc, perhaps near very abrupt disc features.

\subsection{Connections with semiconvection}

Of course, the parasitic theory of saturation may only be part of
the story. While it should reliably predict the initial amplitude of the turbulent flow, 
on longer times the turbulence could evolve according to
other dynamics altogether. The saturation of
semi-convection
is illuminating in
this respect, as it
shares many of the features of convective overstability in discs. 
(Indeed, in two dimensions the mathematical formalisms are almost identical.)
Instead of relying on angular momentum and entropy gradients, 
semiconvection
emerges in the presence of composition and entropy gradients (Kato
1966, Rosenblum et al.~2011). Its nonlinear development takes one of two
courses: (a)
mild turbulent convection or (b)
large-scale `layering' of convective zones over strongly
stratified interfaces (Turner 1968, Merryfield 1995). 
The turbulent transport in the second course is far greater than in
the first by at least an order of magnitude, 
but it only arises when the background stratification is
sufficiently strong (Rosenblum
et al.~2011, Mirouh et al.~2012). In fact, there exists a critical buoyancy
frequency $|N^2_c|$,
below which the turbulence is weak and above which it suddenly becomes
much more intense. This critical value depends closely on Pr (Mirouh
et al.~2012). 
 
Can the convective overstability in discs 
exhibit similar bimodal behaviour? In the disc context the layers would
be combined zonal and elevator flows, and thus would correspond to
 rings of concentrated vorticity. Indeed, the 2D
simulations of Lyra (2014) develop large-scale radial structure at
late times that could be identified as a `semi-convective layer'. 
Key questions are: what is the critical value of $N^2$
below which radial layers fail to develop (if such a value exists)? Which side of 
this value do realistic dead zones fall? What about other more abrupt
disc features? Obviously if layers dominate
the overstability's long term saturation then
the `parasitic theory' must be discarded, 
and the convective
overstability may instigate more vigorous, and interesting, dynamics. 
Simulations are currently underway to
test these competing ideas.

Zonal flows may be
susceptible to the Kelvin-Helmholtz instability, shedding
vortices as they degenerate. Lyra's 3D simulations also
 manifest vortices, though it is unclear if they arise from the
 breakdown of zonal layers or by the SBI, seeded
 by vigorous non-axisymmetric turbulence. Again it is important to
 note that these simulations are strongly stratified, with $n^2\approx
 0.1$. More realistic values may yield a less vigorous and more
 axisymmetric state, one that may struggle to seed vortices directly. 
Again this needs to be checked in dedicated 3D simulations.

\section{Discussion}

Ordinarily the angular momentum gradient in a
 protoplanetary disc is sufficiently strong to
 stabilise a negative entropy gradient, if one exists.
But because thermal diffusion is far more efficient than
viscous diffusion, double diffusive instabilities arise that
can unleash the energy stored in the adverse gradient. These include
the subcritical
baroclinic instability and the convective overstability 
(Lesur \& Papaloizou 2010, Klahr
\& Hubbard 2014, Lyra 2014). The resistive
instability, on the other hand, uses magnetic fields to diffuse angular
momentum faster than heat and is hence double-diffusive in the
opposite sense (Latter et al.~2010a). These various
mechanisms may liven up the dead zones of protoplanetary
discs, though it is unclear if any one of them is the answer to the
question of angular momentum transport.

In this paper I revisit the linear and nonlinear dynamics of the
convective overstability. Because the linear modes are also nonlinear
solutions, my focus has been on the parasitic modes that limit their
amplitude, the idea being that the parasites control the
saturation level of the ensuing turbulent dynamics. This approach predicts
that the overstability generates only very weak turbulence, with a
strain field
$\sim |N^2|/\Omega \sim 10^{-3}\Omega$. The conclusion is that the
flow remains axisymmetric and rather gentle, unable to transport much
angular momentum nor generate vortices. However, near very abrupt disc
structures, such as edges, $|N^2|$ may be larger and greater activity
might be anticipated.
But I also explore alternative
ideas, drawing on recent work in semi-convection that shows when
$|N^2|$ crosses a critical value the flow splits into
thermo-compositional layers that greatly enhance transport (Rosenblum et
al.~2011, Mirouh et al.~2012). Something similar may occur in
simulations of the overstability, the layers taking the form of `zonal
flows' (Lyra 2014). Future work should establish the critical $|N^2|$
above which this takes place, and whether we expect this
behaviour in realistic discs. 
 
If convective overstability is present, which is not always assured,
what is its role in the disc dynamics?  
One possible application is 
to the excitation of random motions in disc solids, thereby
influencing their collision speeds and frequencies. The greatest
effect will be on marginally coupled particles, whose stopping
time is similar to the typical turbulent turnover time ($\sim
1/\Omega$ for inertial wave turbulence). These particles should have
radii of roughly 10 cm to 1 m (Chiang \& Youdin 2010). The
velocity dispersion induced by the coupling will be of order
$v_\text{turb}$, a fairly mild enhancement in most cases.
The turbulent flow may also concentrate such particles, further
amplifying collision rates, though this can only be checked by 
detailed numerical simulations (Hogan \& Cuzzi 2007, Pan \& Padoan
2013). 

In conclusion, the significance of the overstability, vis-a-vis other hydrodynamical
processes, essentially comes down to the magnitude of $N^2$. It
sets the base level of turbulent motions (cf.\ Eq.~\eqref{turb}) and
whether the system selects a gentle or more vigorous state (cf.\
Section 5.2). Thus further constraints on protoplanetary disc structure and
further numerical simulations are needed to help properly assess the
instability's place in disc dynamics.

\section*{Acknowledgments}
I thank the reviewer, Steve Balbus, for a helpful set of comments that
improved the manuscript.
I also thank Gordon Ogilvie, 
Sebastien Fromang, Geoffroy Lesur, Andrew Youdin, and John Papaloizou
for helpful tips and suggestions. I am also grateful to Hubert Klahr and Wlad Lyra for
clarifying some of their work and for inspiring me to have a look at
the problem. Finally I am indebted to Jerome Guilet who generously
read through an earlier version of the manuscript and who suggested
valuable improvements to Section 4.4 particularly. 
This research is partially funded by STFC grant
ST/L000636/1.

\appendix

\section{Viscous linear theory}
If viscosity is introduced into the linearised equations of
\eqref{lin1}-\eqref{lin3} one can derive the following dispersion
relation
\begin{equation}
s^3 + a_2 s^2 + a_1 s + a_0 =0,
\end{equation}
where 
\begin{align*}
a_2 &= K^2\xi + 2 K^2 \nu, \\
a_1 &= K^4\nu^2 + 2 K^4 \xi\nu + \kappa^2 + N^2, \\
a_0 &= K^2\xi( K^4 \nu^2 + \kappa^2) + K^2 \nu N^2,
\end{align*}
(see also Guilet \& M\"ulller 2015). 

In the regime $n^2\gg \text{Pr}$ the viscosity only adds
an order Pr correction to the maximum growth rate. On smaller scales
its effects are more severe and it will ultimately stabilise the
modes. I first derive the critical $K$ when
this occurs. As this corresponds to a Hopf bifurcation,
$s=\text{i}\omega$ at criticality, in which $\omega$ is a real frequency.
On substituting this into the dispersion relation, one obtains the two
equations
$$ -\omega^3 + a_1 \omega=0, \qquad -a_2\omega^2 + a_0 =0,$$
which, after some manipulation, yield an expression for the critical wavenumber
\begin{equation}\label{kcrit}
K^4_c = -\frac{2\text{Pr}- n^2(1+\text{Pr})}{2\text{Pr}(1+\text{Pr})^2}\frac{\xi^2}{\kappa^2}.
\end{equation}
For small Pr, we have the useful estimate
\begin{equation}
K_c \approx \left(\frac{n^2}{\text{Pr}}\right)^{1/4}K_\text{fast},
\end{equation}
where $K_\text{fast}\approx \sqrt{\xi/\kappa}$ 
is the wavenumber of the fastest growing mode.
Note in particular the quartic root. Though both Pr and $n^2$ are
small, the viscous cutoff need not be so well separated from
$K_\text{fast}$. If $n^2\sim 10^{-3}$ and Pr $\sim 10^{-7}$ then
$K_c\sim 10 K_\text{fast}$. The unusual quartic expression comes
about because the viscous frequency $\sim \nu K^2$ and
the growth rate at small $K$ is $\sim -N^2/(\xi K^2)$, from
Eq.~\eqref{sapprox}. Equating the two frequencies gives rise to that
fourth power.

In an inviscid gas all scales are unstable when $N^2<0$. 
Viscosity, however, alters this condition.
It kills off the instability everywhere
whenever the numerator in
\eqref{kcrit} is negative. This yields the following instability criterion
\begin{equation}\label{visccrit}
N^2 < -\frac{2\text{Pr}}{1+\text{Pr}}\Omega^2,
\end{equation}
which replaces that of Schwarzchild. Thus the entropy gradient must be
both negative \emph{and} sufficiently strong. According to Section
3.4, the right side of \eqref{visccrit} is $\sim 10^{-7}$, and so in
practice only
exceptionally weak negative gradients are stabilised. In fact,
we need to also consider the outer scale: 
$K_c> 1/H$ or else the instability will not fit into the disc. The
instability criterion then picks up another small correction term, and becomes
\begin{equation}
N^2 <  -\frac{2\text{Pr}}{1+\text{Pr}}\Omega^2 -\frac{2\text{Pr}(1+\text{Pr})}{\text{Pe}^2}.
\end{equation}

\section{Parametric instability}

In this section I solve the system \eqref{LIN1}-\eqref{LIN2} to
leading order in the small oscillation amplitude, $\epsilon$. From the
start I set $m=0$, to make life simple.

The followings expansions are assumed:
$\sigma = \epsilon\sigma_1 + \dots$, and
\begin{align*}
\hat{u}_x &= u_0 + \epsilon u_1 + \dots, \qquad
\hat{u}_y = v_0 + \epsilon v_1 + \dots, \\
\hat{u}_z &= w_0 + \epsilon w_1 + \dots, \qquad
\hat{p} = p_0 + \epsilon p_1 + \dots. 
\end{align*}
These are thrown into the linearised equations and the various orders
in $\epsilon$ collected. At leading order the convective oscillation
does not appear, and the four equations can be
combined into $\mathcal{L}p_0 = 0$, where
\begin{equation}
\mathcal{L} = (\d_t^2 + 1)\d_z^2 - k^2 \d_t^2.
\end{equation}
This equation has solutions $p_0 \propto E_n \equiv \ee^{\ii n z - \ii \omega t}$
for vertical wavenumber $n$ and frequency $\omega$. Because the
solutions must be $2\pi$-periodic in $z$, the wavenumber $n$ takes
integer values. For such solutions to exist, however, the following
solvability condition must hold,
\begin{equation} \label{inertial}
\omega^2 = \frac{n^2}{k^2 + n^2},
\end{equation}
which is, of course, the dispersion relation for inertial waves in
accretion discs. The general solution at this order is an infinite sum
(over $n$) of the individual inertial waves. However we are only 
interested in neighbouring modes that can come into resonance with
the background oscillation. These are the $n$ and $n+1$ modes, with
associated frequencies $\omega_1$ and $\omega_2$, calculated 
from Eq.~\eqref{inertial}. For resonance to occur
$\omega_1+\omega_2=1$, which yields the condition
Eq.~\eqref{resonant}, in the main text. Thus
\begin{equation}
p_0 = A E_n + B E_{n+1},
\end{equation}
where $A$ and $B$ are constants. Without loss of
generality, we set $B=1$.
Finally, the velocity components follow from
\begin{align}
u_0= -\frac{n^2}{\omega k}p_0, \quad v_0
=\frac{n^2\ii}{2k\omega^2}p_0, \quad w_0=\frac{n}{\omega}p_0.
\end{align}

At the next order, we obtain $\mathcal{L}p_1 = f$, where the right
hand side is
\begin{equation}
f = (\d_t^2 a + 2 \d_t b)\ii k + (\d_t^2+1)\d_z c,
\end{equation}
and
\begin{align*}
a&=\sigma u_0 + \ii k U\, u_0 + \d_z U\,w_0, \\
b&= \sigma v_0 + \ii k U\,v_0 + \d_z V\, w_0, \\
c&=\sigma w_0 + \ii k U\, w_0.
\end{align*}
Solvability of this equation requires that $f$ possesses no term
proportional to the solution of the homogeneous problem. 
Multiplying $f$ by $E_n^*$ and integrating over $t$ and $z$, then
doing the same with $E_{n+1}^*$, gives two solvability conditions,
essentially equations for $A$ and
$\sigma$. After some tedious algebra, we obtain the
growth rate, as a function of $n$, which is Eq.~\eqref{smallass} in the main
text.

At large $n$, the resonance condition is satisfied when
$$ k^2 = 3n^2 + 3n + \mathcal{O}(1),$$
which correspond to a frequency 
$\omega_1 = -\frac{1}{2} +
\frac{3}{16}n^{-1}+\mathcal{O}(n^{-2})$. Putting these expressions
into \eqref{smallass} gives, to leading order, $\sigma_1=
\frac{3\sqrt{3}}{32}$.

\end{document}